%% file: master.tex
\documentclass[journal]{IEEEtran}

\usepackage{cite}
\usepackage[pdftex]{graphicx}
\graphicspath{{./fig/}}
\DeclareGraphicsExtensions{.pdf,.jpeg,.png,.tikz}

\usepackage[cmex10]{amsmath}
\usepackage[caption=false,font=footnotesize]{subfig}
\usepackage{dblfloatfix}
\usepackage{tipa}
\usepackage{url}
\usepackage{multirow}
\usepackage{tabu}
\usepackage{color}
\usepackage[binary-units]{siunitx}
\newcommand{\bps}{\si{\bit\per\second}}
\newcommand{\kbps}{\si{\kilo\bit\per\second}}
\DeclareSIUnit[number-unit-product = {}]\syllable{syll}

\newcommand{\rv}{\color{black}}
\newcommand{\rf}{\color{black}}

\ifCLASSOPTIONcaptionsoff
 \usepackage[nomarkers]{endfloat}
\let\MYoriglatexcaption\caption
\renewcommand{\caption}[2][\relax]{\MYoriglatexcaption[#2]{#2}}
\fi

\let\MYorigsubfloat\subfloat
\renewcommand{\subfloat}[2][\relax]{\MYorigsubfloat[]{#2}}

\usepackage{tikz}
\usepackage{tikzscale}

\begin{document}
%
\title{Composition of Deep and Spiking Neural Networks for Very Low Bit Rate
  Speech Coding}


\author{Milos~Cernak,~\IEEEmembership{Member,~IEEE,}
        Alexandros~Lazaridis,~\IEEEmembership{Member,~IEEE,}
        Afsaneh~Asaei,~\IEEEmembership{Senior Member,~IEEE,}
        Philip~N.~Garner,~\IEEEmembership{Senior Member,~IEEE}
\thanks{Authors are with Idiap Research Institute, Centre du Parc,
  Rue Marconi 19, 1920 Martigny, Switzerland, contact details:
  http://people.idiap.ch/mcernak.}
}


\maketitle

\begin{abstract}
\input{abstract}
\end{abstract}


\IEEEpeerreviewmaketitle

\section{Introduction}
\label{sec:intro}
\input{introduction}

\section{Neural Networks for Speech Coding}
\label{sec:nn}
\input{networks}

\section{Open-Source Experimental Framework}
\label{sec:framework}
\input{framework}

\section{Evaluation}
\label{sec:experiments}
\input{experiments}

\section{Conclusions}
\label{sec:conclusion}
\input{conclusions}

\section*{Acknowledgment}

This work has been conducted with the support of the Swiss
NSF under grant CRSII2 141903: Spoken Interaction with
Interpretation in Switzerland (SIWIS), and under SP2: the
SCOPES Project on Speech Prosody. The wor was also partly supported
under the RECOD project by armasuisse, the Procurement and Technology
Center of the Swiss Federal Department of Defence, Civil Protection
and Sport.

Afsaneh Asaei has been supported by SNSF
project on ``Parsimonious Hierarchical Automatic Speech Recognition
(PHASER)'' grant agreement number 200021-153507.

\ifCLASSOPTIONcaptionsoff
  \newpage
\fi

\bibliographystyle{IEEEtran}
\bibliography{alaza_refs,refs,refs-older}


\begin{IEEEbiography}[{\includegraphics[width=1in,height=1.25in,clip,keepaspectratio]{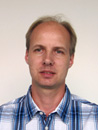}}]{Milos~Cernak}
holds a Ph.D. in Telecommunications from Slovak University of
Technology in Bratislava. Since 2011 he is a senior engineer and
scientific collaborator at Idiap Research Institute, Martigny,
Switzerland, involved in various  Swiss R\&D academic and industrial
projects with a core focus on speech recognition and synthesis, low
bit rate speech coding, and prosody parametrization. From 2008, he was
a member of IBM Research in Prague, working on embedded ViaVoice
speech recognition and IBM Reading Companion. After graduating in
2005, he was a post-doc researcher at Institute EURECOM in France, and
a principal researcher at Slovak Academy of Sciences. During his
studies in 2001, he was also a visiting scientist at Iowa State
University's Virtual Reality Application Centre, Ames, USA. He is
interested in signal processing, phonology, and neural basis of speech
production and perception. He was a plenary lecturer for the 2016
Frederick Jelinek Memorial Summer Workshop.
\end{IEEEbiography}

\begin{IEEEbiography}[{\includegraphics[width=1in,height=1.25in,clip,keepaspectratio]{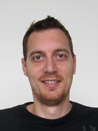}}]{Alexandros~Lazaridis} 
was born in Thessaloniki, Greece, in 1981. He graduated in 2005
(Diploma) from the Department of Electrical and Computer Engineering of the
Aristotle University of Thessaloniki, Greece. He received his PhD
degree in Feb. 2011 from the Department of Electrical and Computer
Engineering of the University of Patras, Greece. Since Nov. 2012 he is
a post-doctoral researcher at Idiap Research Institute, Martigny,
Switzerland. He is author and co-author in more than 25 publications
in scientific journals and international conferences. His research
interests include speech and audio signal processing, speech
synthesis, speech prosody and spoken language/dialect/accent
identification.
\end{IEEEbiography}

\begin{IEEEbiography}[{\includegraphics[width=1.1in,height=1.25in,clip,keepaspectratio]{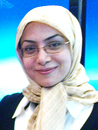}}]{Afsaneh
    Asaei} received the B.S. degree from Amirkabir University of
  Technology and the M.S. (honors) degree from Sharif University of
  Technology, in Electrical and Computer engineering,
  respectively. She held a research engineer position at Iran
  Telecommunication Research Center (ITRC) from 2002 to 2008. She then
  joined Idiap Research Institute in Martigny, Switzerland as a
  doctoral research assistant and Marie Curie fellow on speech
  communication with adaptive learning training network. She received
  the Ph.D. degree in 2013 from \'Ecole Polytechnique F\'ed\'erale de
  Lausanne. Her thesis focused on structured sparsity for multiparty
  reverberant speech processing, and its key idea was awarded the IEEE
  Spoken Language Processing Grant.
Currently, she is a postdoctoral researcher at Idiap Research
Institute. She has served as a guest editor of Speech Communication
special issue on Advances in sparse modeling and low-rank modeling for
speech processing and co-organized special issues on this topic at
HSCMA'2014 and LVA/ICA'2015. Her research interests lie in the areas
of signal processing, machine learning, statistics, acoustics,
auditory scene analysis and cognition, and sparse signal recovery and
acquisition. 
\end{IEEEbiography}

\begin{IEEEbiography}[{\includegraphics[width=1in,height=1.25in,clip,keepaspectratio]{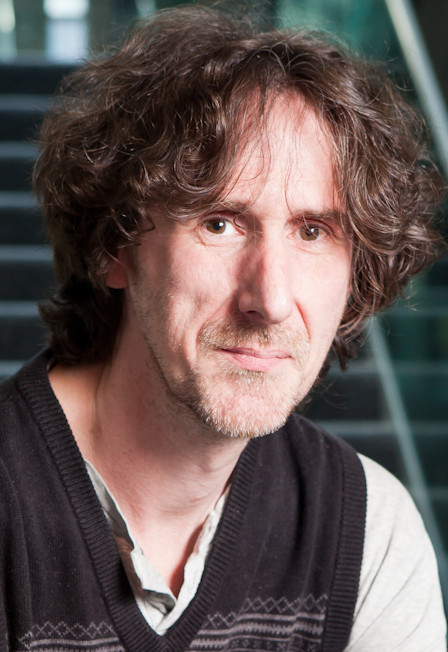}}]{Philip~N.~Garner}
received the degree of M.Eng. in Electronic Engineering from the
University of Southampton, U.K., in 1991, and the degree of Ph.D. (by
publication) from the University of East Anglia, U.K., in 2012. He
first joined the Royal Signals and Radar Establishment in Malvern,
Worcestershire working on pattern recognition and later speech
processing. In 1998 he moved to Canon Research Centre Europe in
Guildford, Surrey, where he designed speech recognition metadata for
retrieval. In 2001, he was seconded (and subsequently transfered) to
the speech group at Canon Inc. in Tokyo, Japan, to work on
multilingual aspects of speech recognition and noise robustness. As of
April 2007, he is a senior research scientist at Idiap Research
Institute, Martigny, Switzerland, where he continues to work in
R\&D of speech recognition, synthesis and signal
processing. He is a senior member of the IEEE, and has published
internationally in conference proceedings, patent, journal and book
form as well as serving as coordinating editor of ISO/IEC 15938-4
(MPEG-7 Audio).
\end{IEEEbiography}









\end{document}

%% file: abstract.tex
Most current very low bit rate (VLBR) speech coding systems use hidden Markov
model (HMM) based speech recognition and synthesis techniques. This allows
transmission of information (such as phonemes) segment by segment; this
decreases the bit rate. However, an encoder based on a phoneme speech
recognition may create bursts of segmental errors; these would be further
propagated to any suprasegmental (such as syllable) information
coding. Together with the errors of voicing detection in pitch parametrization,
HMM-based speech coding leads to speech discontinuities and unnatural speech
sound artefacts.

In this paper, we propose a novel VLBR speech coding framework based on neural
networks (NNs) for end-to-end speech analysis and synthesis without HMMs. The
speech coding framework relies on a phonological (sub-phonetic) representation
of speech. It is designed as a composition of deep and spiking NNs: a bank of
phonological analysers at the transmitter, and a phonological synthesizer at
the receiver.  These are both realised as deep NNs, along with a spiking NN as
an incremental and robust encoder of syllable boundaries for coding of
continuous fundamental frequency (F0). A combination of phonological features
defines much more sound patterns than phonetic features defined by HMM-based
speech coders; this finer analysis/synthesis code contributes to smoother
encoded speech. Listeners significantly prefer the NN-based approach due to
fewer discontinuities and speech artefacts of the encoded speech. A single
forward pass is required during the speech encoding and decoding. The proposed
VLBR speech coding operates at {\rv a bit rate of approximately} 360 bits/s.

%% file: introduction.tex
The ITU-T standardisation effort for speech coders operating below
{\rv \num{4000} bits per second (\bps)} began in 1994 \cite{Dimolitsas1994}.
However, it has been shown to be difficult to achieve toll-quality
performance in all conditions, such as intelligibility, quality,
speaker recognizability, communicability, language independence and
complexity. All these conditions are exposed even more in speech
coders operating at VLBR speech coding of the order of hundreds of \bps.

To achieve VLBR of \numrange{200}{500} \bps, parametric speech coding
based on a recognition/synthesis paradigm has been proposed. Approaches
following this paradigm can be classified into two categories:
corpus-based, e.g., \cite{Lee2001,Baudoin03}, and hidden Markov model
(HMM) based, e.g., \cite{Picone89,Tokuda98,McCree2008}. This
classification follows trends in speech synthesis, where popular
unit-selection methods are replaced by HMM-based parametric speech
synthesis methods. The 
parametric methods benefit from better adaptation properties and lower
footprint. However, most current HMM-based VLBR systems have complex
designs. The phonetic encoder --- automatic speech recognition (ASR) ---
consists of acoustic HMMs and language models and an incremental
search module.  Similarly, the phonetic decoder requires acoustic
HMMs, including a streaming/performative HMM-based speech synthesis
system and an incremental speech vocoder.

Our recent work~\cite{Cernak15ieee} also focused on HMM-based VLBR
speech coding, designing a coder operating with an acceptable
communication delay for real-time speech communication, and with a view to
be exploited in military and tactical communication systems. 

Closer analysis of our HMM-based encoder --- a phoneme ASR system --- revealed
that it sometimes created bursts of segmental errors when the recognition
failed. This is a well known phenomenon; phoneme ASR misrecognition rates tend
to increase with longer words~\cite{greenberg00:linguistic}. In addition, the
coder~~\cite{Cernak15ieee} detected syllable boundaries from recognised phoneme
sequences based on the sonority sequencing principle; thus phoneme
misrecognition was further propagated to the syllable boundary estimation
sometimes resulting in wrongly detected syllables. The HMM coder also used
voiced/unvoiced detection for parametrization of the F0 signal in the voiced
regions. This F0 encoding, plugged-in to the VLBR system, created additional
speech discontinuities and unnatural speech sound artefacts. Moreover,
canonical phone indexes transmitted from a transmitter to a receiver compressed
all phonetic variability to the order of tens of phonetic
categories. Increasing the number of categories increased the bit rate.

In this paper, we aim to address the above limitations of {\rv such
speech codecs for very low bit rates that are composed of HMM-based
recognition/synthesis components}. First, we propose to
\textit{replace HMMs by deep NNs} {\rv that perform} speech analysis
and synthesis based on a phonological speech representation. The
phonological speech representation extends the encoded phonetic
variability to the order of hundreds or even thousands of sound
categories.
Second, we propose to use \textit{a spiking NN} as a neuromorphic,
incremental, and highly noise-robust syllable boundary detector, used
for syllable-based continuous F0 signal coding. Using continuous F0
modelling should also alleviate speech re-synthesis discontinuities
caused by erroneous detection of unvoiced segments. This end-to-end NN based
speech coding (called \textit{NN speech coder} hereinafter) should
significantly reduce the current design (and hopefully also computational)
complexity since the speech encoding and decoding would be realised as a
single NN forward pass.





In this work, we consider the three phonological systems defined in Appendix A
of~\cite{Cernak2016a}: (i) the Government Phonology
(GP)~\cite{Harris94,Harris95}, (ii) the Sound Pattern of English
(SPE)~\cite{chomsky68sound} and (iii) the extended SPE system
(eSPE)~\cite{Yu2012,Siniscalchi2012}. Each phoneme is represented by its
sub-phonetic attributes, or phonological classes. A vector of all phonological
class probabilities is referred to as {\rv the} phonological posterior. There
are only very few phonological classes comprising a short term speech signal;
hence, the phonological posterior is a sparse vector, allowing very few
combinations that can be stored in a codebook for VLBR speech
coding~\cite{Asaei15compr}. The temporal span of phonological features is wider
than the span of phonetic features, and thus the frame shift can be higher,
i.e., fewer frames are transmitted yielding lower bit rates.

The rest of the paper is organized as follows. Section \ref{sec:nn}
contains a review of usage of NNs for speech coding. In section
\ref{sec:framework} an open-source experimental framework
used in this work is introduced, and in Section \ref{sec:experiments} the
results are presented of a comparison of HMM-based and NN-based VLBR speech
coders. Finally, conclusions are drawn in Section \ref{sec:conclusion}.

%% file: networks.tex
\subsection{Background}

Recently, in the speech recognition/analysis field, a shift has been
observed from HMM-based approaches towards the use of deep neural
networks (DNNs)~\cite{Bengio2009}. Even though the concept of using neural networks in
recognition is not new \cite{Bourlard1994}, the increase of
available speech resources and of computational power, and the use of
graphics processing units, have led to a major research interest in
DNNs. Additionally, deep architectures with multiple layers can
overcome the limitations in the representational capability of HMMs,
which are incapable of modelling multiple interacting source streams
\cite{Mohamed2009}. Furthermore, DNNs are able to create non-linear
mappings between the input and output features which cannot be
achieved by using Gaussian mixture models (GMMs) in HMM-based
approaches, making them more appropriate for modelling the speech 
signal~\cite{Hinton2012}. As a result, DNN-based approaches have
{\rv shown} superior performance in comparison to HMM-based ones
in recognition tasks \cite{Dahl2012,Hinton2012}.

The same shift has also been observed in the text-to-speech (TTS) field. There
are various limitations and drawbacks which occur in HMM-based
TTS~\cite{Zen2013}, e.g., inefficiency to express complex dependencies in the
feature space~\cite{Esmeir2007}; this leads to decision trees becoming
exceedingly large, inefficient and data hungry. Also, the use of decision trees
leads to the fragmentation of the training data, linking specific parts of them
to each terminal node of the tree~\cite{Yu2011}. These limitations led to the
introduction of DNNs in the field of parametric speech synthesis as well,
constantly outperforming HMM-based speech synthesis
systems~\cite{Zen2013,Qian2014,Lazaridis2015a}.

In the context of speech coding, NNs are used in
\textit{waveform-approximating coders}, a family of coders originated
in~\cite{Atal82,schroeder1985}, to address either improving quality or
reducing computational complexity. The former usage aims to {\rv replace
linear prediction of speech samples or parameters of the excitation
signal with a non-linear prediction. The non-linear prediction is
usually based on multilayer perceptrons}~\cite{Morishima90,Zhen96,Hunt96,Chavy99,Fz1999}. Recently, 
regression-based packet loss concealment was proposed using
DNNs~\cite{Lee16}. The latter usage aims to reduce the complexity of
the codebook search process or gain prediction~\cite{Sheikhan09} using,
for example, recurrent NNs~\cite{Easton91,Wu94}.

Speech coding based on speech modelling is known as \textit{parametric coding},
where the parameters of the speech models are transmitted, as in, e.g., a
multiband excitation vocoder~\cite{Griffin88}. As in waveform coding, a
multilayer perceptron was proposed to decrease the computation complexity of
the codebook of line spectral frequencies in the 800 {\rv \bps} multiband
excitation speech coding~\cite{Cui98}.

While NNs have been used in previous speech coding approaches only as isolated
modules targeting some particular computation, we propose \textit{end-to-end
  VLBR NN-based coder}, aiming to replace HMM-based speech analysis and
synthesis by deep and spiking NNs.


\subsection{Coding of phonetic and phonological information}
\label{sec:segmental}

VLBR coding has its roots in the linear prediction
model~\cite{Wong82,Roucos82,Roucos83,Wong83,Tsao85,Shiraki88}, as do
the majority of standardised higher bit-rate speech
codecs~\cite{Gibson16}. Unlike such coding, which aims to compress the
parameters of the linear prediction model, we aim to keep the
parameters of the linear model.  Rather, we move the compression one
level above, from the acoustics to the phonetic and phonological
representation of the speech signal. {\rf Our previous
  work~\cite{Cernak15ieee} showed that, if such a representation is
  inferred from the acoustics well, we could decrease the transmission
  rate to about 200 \bps, however the speech quality did not achieve
  the quality of old LPC coders operating at 500 \bps~or less, that is,
  slightly lower quality than normal LPC coding at 2 \kbps. In this
  work, we hypothesise that recent advances in the convergence of deep
  learning and speech technology could fill the gap, i.e.,
  improving the speech quality of VLBR coding whilst still using a phonetic
  and phonological speech representation. It is also fair to note
  that using the phonetic and phonological speech representation is
  still language dependent, particularly in speech synthesis.} 

Encoding of segmental information starts with analysis by converting a
segment of speech samples into a sequence of acoustic features
$X=\{\vec{x}_1,\ldots,\vec{x}_n,\ldots,\vec{x}_N\}$, where $N$ denotes
the number of segments in the utterance. Conventional cepstral
coefficients can be used as acoustic features. Encoding can be done on
a phonetic or phonological (sub-phonetic) level.

In the former case, the acoustic feature observation  sequence $X$ is
converted into a parameter sequence
$\vec{z}_n=[z_n^1,\ldots,z_n^p,\ldots,z_n^P]^\top$,
where the $n$-th frame consists of posterior probabilities
$z_n^p=p(c_p|x_n)$ of $P$ classes (phonemes), and $.^\top$  is
the transpose operator. The a-posteriori estimates $p(c_p|x_n)$ are $0
\le p(c_p|x_n) \le 1, \forall p$ and $\sum_{p=1}^P p(c_p|x_n) =
1$. All the phonemes have to be recognised to access higher semantic
levels (words and utterances); hence, using the phone posterior
probabilities can be considered a sequential scheme. The phonetic
vocoding, where only one encoded segment (a
phoneme) is transmitted each time, is an example.

In the latter case, the acoustic feature observation  sequence $X$ is
converted into a parameter sequence
$\vec{z}_n=[p(c_1|x_n),\hdots,p(c_k|x_n),\hdots,p(c_K|x_n)]^\top$, that
consists of $K$ phonological class-conditional posterior
probabilities, where $c_k$ denotes the phonological class. The
phonological posteriors are computed by a bank of parallel DNNs, each
estimating the posteriors $z_n^k$ as probabilities that the $k$-th
phonological feature occurs (versus does not occur). The a-posteriori
estimates $p(c_k|x_n)$ are also $0 \le p(c_k|x_n) \le 1, \forall k$,
but $\max \sum_{k=1}^K p(c_k|x_n) = K$. Only very few classes are
active during a short term signal, $\sum_{k=1}^K p(c_k|x_n) \ll K$,
resulting in a sparse vector $\vec{z}_n$. Using the phonological
posterior probabilities can be considered a parallel scheme via $K$
different phonological classes. 

While erroneous phone posterior
estimation leads to a possible failure of the higher semantic segment
recognition, erroneous phonological posterior estimation leads to a
failure only at a sub-phonetic feature level, and this partial
error does not necessarily lead to misrecognition of the whole
recognized segment.




Decoding of segmental information is realised as a DNN that learns the
highly-complex mapping of the parameter sequence, $Z$, to the speech
parameters~\cite{Cernak15icassp}. It consists of two computational
steps. The first step is a DNN forward pass that generates the speech
parameters (the LPC speech parameters described in
Section~\ref{sec:cepstraldnn}); the second one is generation of
the speech samples from the speech parameters.

\subsection{Coding of prosodic information}

Speech analysis results in discrete units while moving from segmental
to suprasegmental (prosodic) level of speech representation.
This discrete information can be estimated directly from the speech
signal. For example, the Probabilistic Amplitude Demodulation method
proposed in \cite{Turner11} can robustly estimate the syllable and
stress amplitude modulations using a representation of
electrophysiological recordings of the auditory cortex. The work of
\cite{Leong14jasa} proves that phase relations of the amplitude
modulations, known as hierarchical phase locking and nesting, or
synchronization across different temporal
granularity~\cite{Lakatos05}, is a good indication of the syllable
stress.

Phonological posteriors have several interesting properties. Even
though they are segmental features by definition, they convey
prosodic information about lexical stress and prosodic accent; this is
embedded in their support (index) of active
coefficients~\cite{Cernak2016b}. In the context of parametric speech
coding it could be interpreted that we do not need to encode this kind of
prosodic information explicitly. Rather, in our current
approach, we focus just on coding of the continuous F0
signal. Modelling continuous F0 has been shown to be more effective in
achieving natural synthesised speech~\cite{Yu10,Yu11}, and can be
effectively used with noisy speech~\cite{Ogbureke12}. We therefore hypothesise
that continuous F0 modelling could improve recognition/synthesis VLBR
coding as well.

Effective encoding of the F0 signal can be realised by curve fitting
done on a syllable level. We thus propose to encode the continuous F0
signal using the discrete (Legendre) orthogonal polynomial (DLOP),
as in~\cite{Cernak15ieee}.
To estimate syllable boundaries from the speech signal, a
neuromorphic oscillatory device is used, based on modelling brain neural
oscillations at syllable frequency.  This results in highly noise robust
incremental syllable boundary detection~\cite{Cernak15neuro}. It is
built around an interconnected network composed of 10 excitatory and 10
inhibitory leaky integrate-and-fire neurons. The spiking NN declares a putative syllable boundary for each
inhibitory spike burst. We selected this approach to alleviate
segmental error propagation in the suprasegmental information coding,
and also keep an end-to-end neural network design for the VLBR system.

In the original proposal of syllable-based F0 parametrization for
speech coding~\cite{Cernak13is}, unvoiced syllables were not
parametrized (and not transmitted); also the pitch coding operated at a
very low 40--60 {\rv \bps}. To achieve similar transmission rates with
continuous F0 coding, we further linearly quantized the DLOP
parameters. Thus, 3-bit quantized second order DLOP (linear) F0 signal
stylization is used for coding of the original F0 signal.

\subsection{Transmission scheme}

Segmental features --- phonological posteriors --- have values mostly
concentrated very close to either 1 or 0, and these binary patterns
allow very efficient use of 1-bit quantization: the probabilities
above 0.5 are normalized to 1 and the probabilities less than 0.5 are
forced to zero.

\begin{figure}[ht]
\centering

\subfloat[Speech samples]{
  \includegraphics[width=\linewidth]{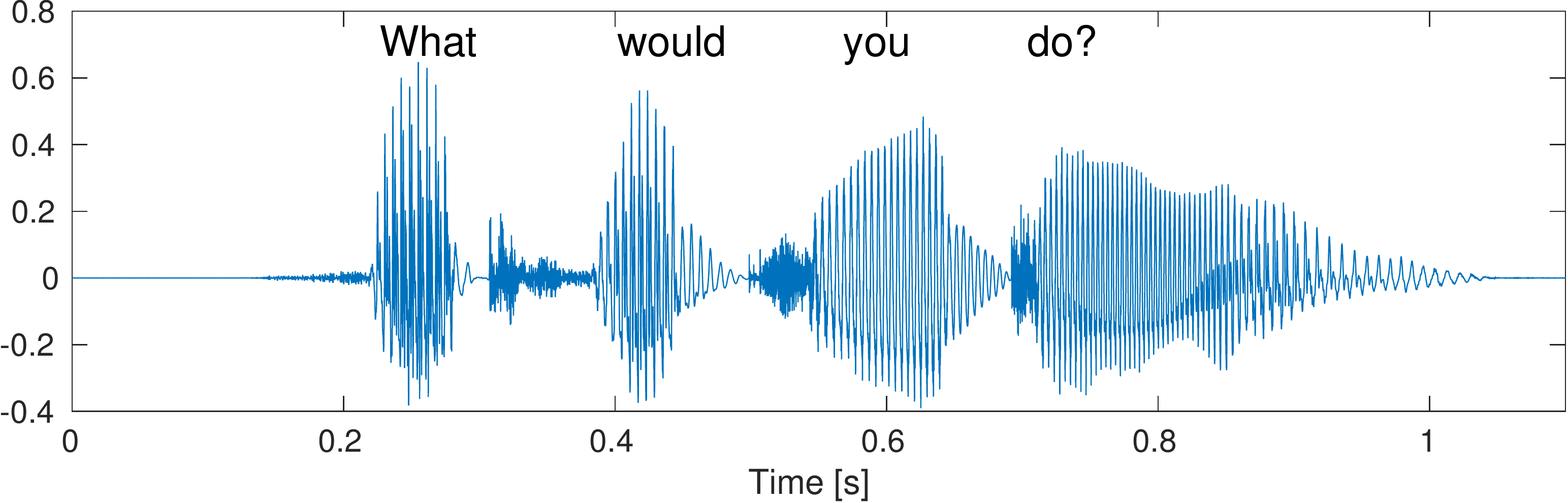} %
\label{fig:ss}
}
\hfil
\subfloat[GP transmission code] {
  \includegraphics[width=\linewidth]{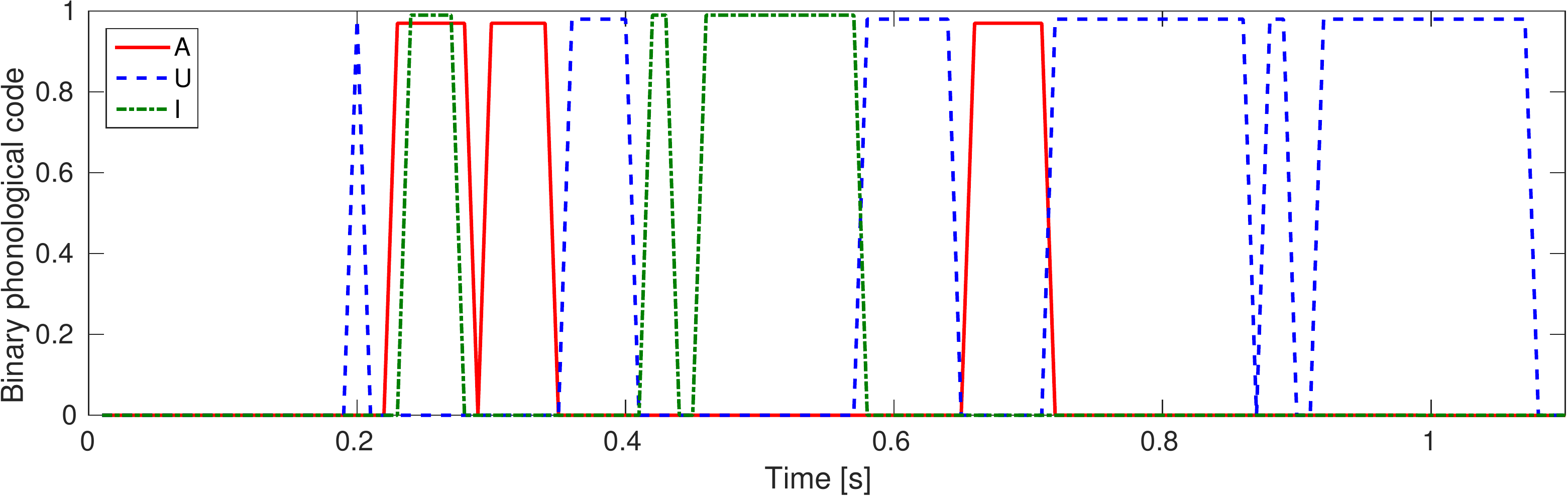} %
\label{fig:gp}
}
\hfil
\subfloat[Pitch transmission code] {
  \includegraphics[width=\linewidth]{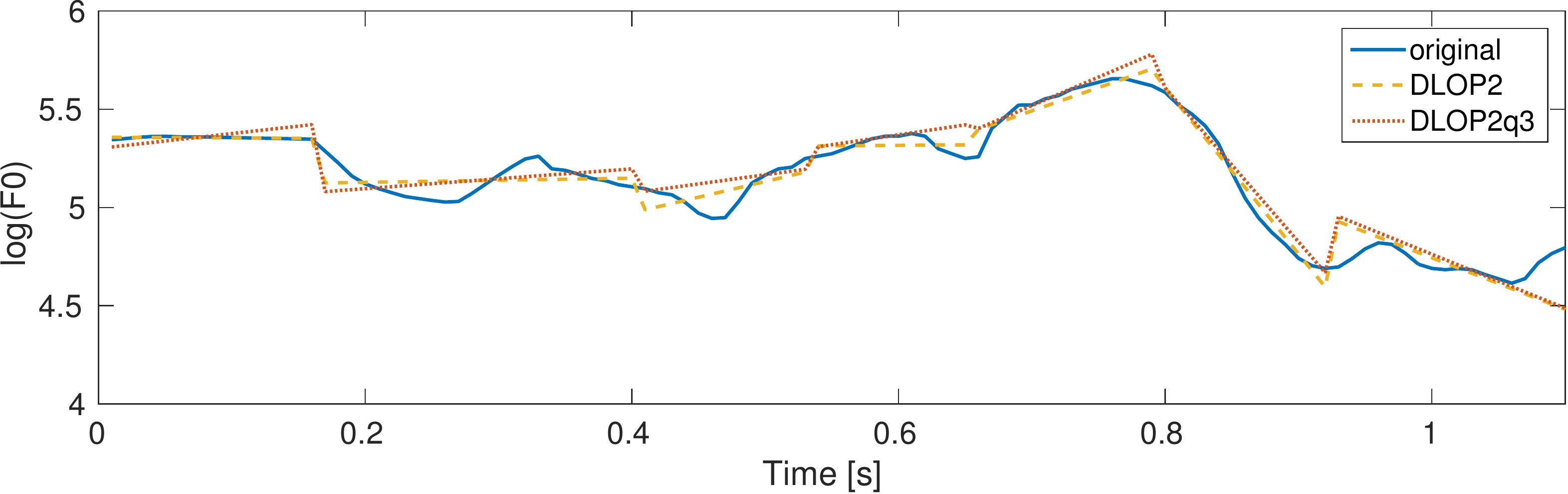} %
\label{fig:pitch}
}
\caption{An illustration of transmission scheme of a demonstration
  sample shown on \ref{fig:ss}, composed of the
  segmental information shown on \ref{fig:gp}: the binary
  phonological code, and prosodic information shown on
  \ref{fig:pitch}: the syllable-based 3bit-quantized second order DLOP
  parameters.}
\label{fig:ex}
\end{figure}

Figure~\ref{fig:ex} illustrates a demonstration sample of the
transmission scheme. Figure~\ref{fig:ss} shows the speech signal.
Figure~\ref{fig:gp} shows binary values of the three basic resonance
phonological primes of the GP system commonly labelled as A, U, I,
denoting the peripheral vowel qualities [a], [u] and [i]
respectively. Other vowels are defined by a composition of the basic
ones; for example, [e] results from fusing the I and A primes. In addition to
these `vocalic' primes, GP also proposes the ``consonantal'' primes that
are omitted in the picture for simplicity.

Figure~\ref{fig:pitch}
shows an original continuous log F0 signal, and linear curve fitting
``DLOP2'' using the syllable boundaries from the SNN. Each syllable is
parametrized by 2 floating point numbers, that are further quantized using
linear 3-bit codebooks, drawn as ``DLOP2q3''. 6 bits are thus needed to
parametrize the first DLOP parameter (the F0 mean) and the second DLOP
parameter (the F0 slope) of the transmitted syllable-based
prosodic code.

%% file: framework.tex
\subsection{Composition of neural networks}

\begin{figure}[ht]
\centering

\subfloat[Encoder]{
\resizebox{3.5in}{!}{%
\begin{tikzpicture}[font=\scriptsize]
  \draw[rounded corners, thick] (0,0.9) rectangle (1,1.6);
  \node[align=center] at (0.5,1.25) {Speech\\Signal};

  \draw[rounded corners, thick] (1.5,1.5) rectangle (3,2.5);
  \node[align=center] at (2.25,2) {Acoustic\\Feature\\Extraction};
  \draw[thick, ->] (0.5,1.6) -- (0.5,2) -- (1.5,2);
  \draw[thick, ->] (0.5,0.9) -- (0.5,0.5) -- (1.5,0.5);

  \draw[rounded corners, thick] (1.50,0) rectangle (3,1);
  \node[align=center] at (2.25,0.5) {Continuous\\Pitch\\Extraction};

  \draw[rounded corners, thick, fill=lightgray] (3.5,0) rectangle (5,1);
  \node[align=center] at (4.25,0.5) {Prosodic\\Analysis\\Spiking NN};
  \draw[rounded corners, thick, fill=lightgray] (3.5,1.5) rectangle (5,2.5);
  \node[align=center] at (4.25,2) {Phonological\\Analysis\\Deep NNs};
  \draw[thick, ->] (3,0.5) -- (3.5,0.5);
  \draw[thick, ->] (3,2) -- (3.5,2);
  \draw[thick, ->] (5,0.5) -- (5.5,0.5);
  \draw[thick, ->] (5,2) -- (5.5,2);

  \draw[rounded corners, thick] (5.5,1.75) rectangle (7.25,2.25);
  \node[align=center] at (6.375,2) {1-bit quant.};
  \draw[rounded corners, thick] (5.5,0.25) rectangle (7.25,0.75);
  \node[align=center] at (6.375,0.5) {Stylisation};

  \draw[thick, ->] (7.25,0.5) -- (8,0.5);
  \draw[thick, ->] (7.25,2) -- (8,2);
\end{tikzpicture}
}%
\label{fig:ENCdesign}
}
\hfil
\subfloat[Decoder] {
\resizebox{2.9in}{!}{%
\begin{tikzpicture}[font=\scriptsize]
  \draw[thick, ->] (0.75,0.5) -- (1.5,0.5);
  \draw[thick, ->] (0.75,2) -- (1.5,2);

  \draw[rounded corners, thick, fill=lightgray] (1.5,1.5) rectangle (3,2.5);
  \node[align=center] at (2.25,2) {Phonological\\Synthesis\\Deep NN};
  \draw[rounded corners, thick] (1.50,0) rectangle (3,1);
  \node[align=center] at (2.25,0.5) {Continuous\\Pitch\\Synthesis};

  \draw[rounded corners, thick] (3.5,1.5) rectangle (5,2.5);
  \node[align=center] at (4.25,2) {Speech\\Re-synthesis};

  \draw[thick, ->] (3,0.5) -- (4.25,0.5) -- (4.25,1.5);
  \draw[thick, ->] (3,2) -- (3.5,2);
  \draw[thick, ->] (5,2) -- (5.5,2);

  \draw[rounded corners, thick] (5.5,1.75) rectangle (7.25,2.25);
  \node[align=center] at (6.375,2) {Speech Signal};

\end{tikzpicture}
}%
\label{fig:DECdesign}
}
\caption{The composition of the functional components of the NN speech
  coder. Three different NNs, shown in grey, are used: (i)
  {\rv phonological analysis DNNs~\cite{Cernak15icassp}} with 1-bit
  quantization encoding the binary segmental code,  (ii) a spiking NN
  with stylisation encoding continuous F0, and (iii) a synthesis
  DNN~\cite{Cernak15icassp} that decodes the segmental code to the
  speech parameters.} 
\label{fig:design}
\end{figure}
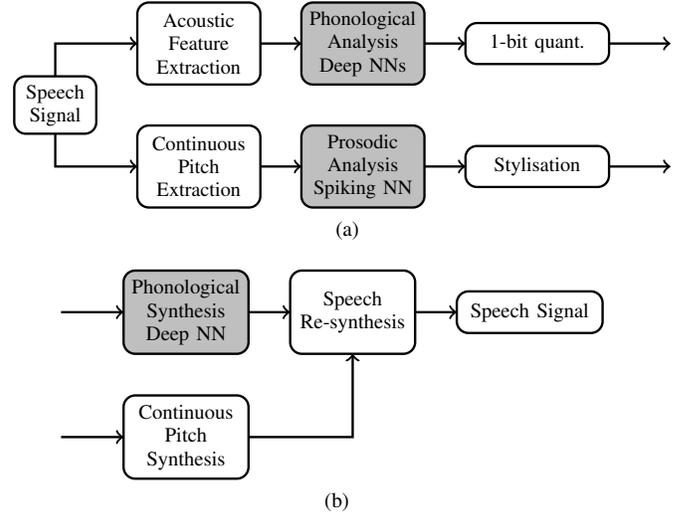

Figure~\ref{fig:design} shows the design of the NN codec. The encoder
shown in Figure~\ref{fig:ENCdesign} is based on a bank of DNNs 
performing segmental speech analysis of conventional acoustic
features, and a parallel spiking NN detecting syllable boundaries of
the continuous F0 signal. The decoder shown in Figure~\ref{fig:DECdesign}
is based on a DNN performing synthesis of speech cepstral parameters
from transmitted segmental and prosodic information.

The outputs of segmental speech analysis are phonological posteriors
where all unique patterns of the training data create the segmental
codebook. The number of unique binary patterns (the size of the
segmental codebook) is a small fraction of the whole permissible
patterns (for example, for the eSPE phonological system, it is about
0.5\%).  The binary patterns are often repeated frame by frame. The
segmental code thus consists of an index of the codebook, along with
the duration of the code. The output of the spiking NN syllable analysis
is stylized using 3-bit quantization of second order DLOP
parameters. All stylized F0 mean and F0 slope values create the
prosodic codebooks, and the prosodic code consists of the two indexes
of the F0 mean and slope codebooks, along with the duration of the
transmitted syllable.

The prosodic code is extracted by spiking NN and DLOP parametrization
independently of the segmental code. That means that the prosodic code
is encoded asynchronously to the segmental one (both codes might have
different start, end, and duration). Both segmental and prosodic codes
are transmitted in parallel.

To confirm the feasibility of the proposed NN speech coder, we
created an experimental framework. {\rf The BSD-licensed open-source
prototype\footnote{\url{https://github.com/idiap/phonvoc/vlbr}},
including pre-trained NNs, is based on:
\begin{itemize}
  \item phonological vocoding performing speech analysis and synthesis
    using phonological posteriors~\cite{Cernak2016}, 
  \item in-house Idiap LPC vocoding, and
  \item syllable onset detection and DLOP-based
    parametrization.
\end{itemize}
}
\subsection{Data}

The DNN encoder is trained on the \textit{si\_tr\_s\_284} set of the
Wall Street Journal (WSJ0 and WSJ1) continuous speech recognition
corpora \cite{WSJDB}, and the SNN is trained on a subset of the TIMIT
corpus~\cite{Consortium1993} (the 10 sentences for speakers indexed
1--100).

The DNN decoder is trained on the Nancy database provided by the Blizzard
Challenge\footnote{\url{http://www.cstr.ed.ac.uk/projects/blizzard/2011/lessac_blizzard2011}}. The
speaker is known as ``Nancy''; she is a US English native female
speaker. The database consisted of 16.6 hours of high quality
recordings of natural expressive human speech recorded in an anechoic
chamber at a 96 kHz sampling rate during 2007 and 2008. The database
comprised of {\rv \num{12095}} utterances, and the following split was
used:
\begin{itemize}
\item the training set, utterances from \num{1} {\rv to \num{10000}},
\item the cross-validation set, utterances from {\rv \num{10001} to
    \num{11000}}, 
\item the test set, the remaining \num{1095} utterances.
\end{itemize}
The text was processed by a conventional and freely available TTS
front-end~\cite{festival}, and the resulting phonetic labels were used
for training of the synthesis DNN. 

The same data was also used to train a baseline HMM-based VLBR speech
coding system.

\subsection{Training}
\label{sec:training}

\subsubsection{Baseline HMM-based system}

We trained the baseline HMM-based system as described
in~\cite{Cernak15ieee}. For HMM analysis models, we trained
three-state, cross-word triphone models with the HTS
variant~\cite{Zen:HTS} of the HTK toolkit on the WSJ training set. We 
tied triphone models with decision tree state clustering based on the
minimum description length (MDL) criterion. The MDL criterion allows
an unsupervised determination of the number of states. In this study,
we obtained 12,685 states each modelled with a GMM consisting of 16
Gaussians. We used mel-frequency cepstral coefficients as acoustic
features. The phoneme set comprising 40 phonemes  (including
``sil'', representing silence) was defined by the CMU pronunciation
dictionary.

For building the HMM synthesis models, the implementation
of training from the EMIME project~\cite{Wester2010} was
used. Five-state, left-to-right, no-skip HSMMs were used. The speech
parameters that were used for training the HSMMs were 39-order
cepstral coefficients, log-F0 and 21-band aperiodicities, along  with
their delta and delta-delta features, framed by \SI{25}{\ms} windows,
extracted every \SI{5}{\ms}. Cepstral instead of mel-cepstral features
were used, as re-synthesis without mel-warping was almost two times
faster.

\subsubsection{Phonological analysis DNNs}

\begin{figure}[ht]
\centering

\subfloat[a]{
  \includegraphics[width=\linewidth]{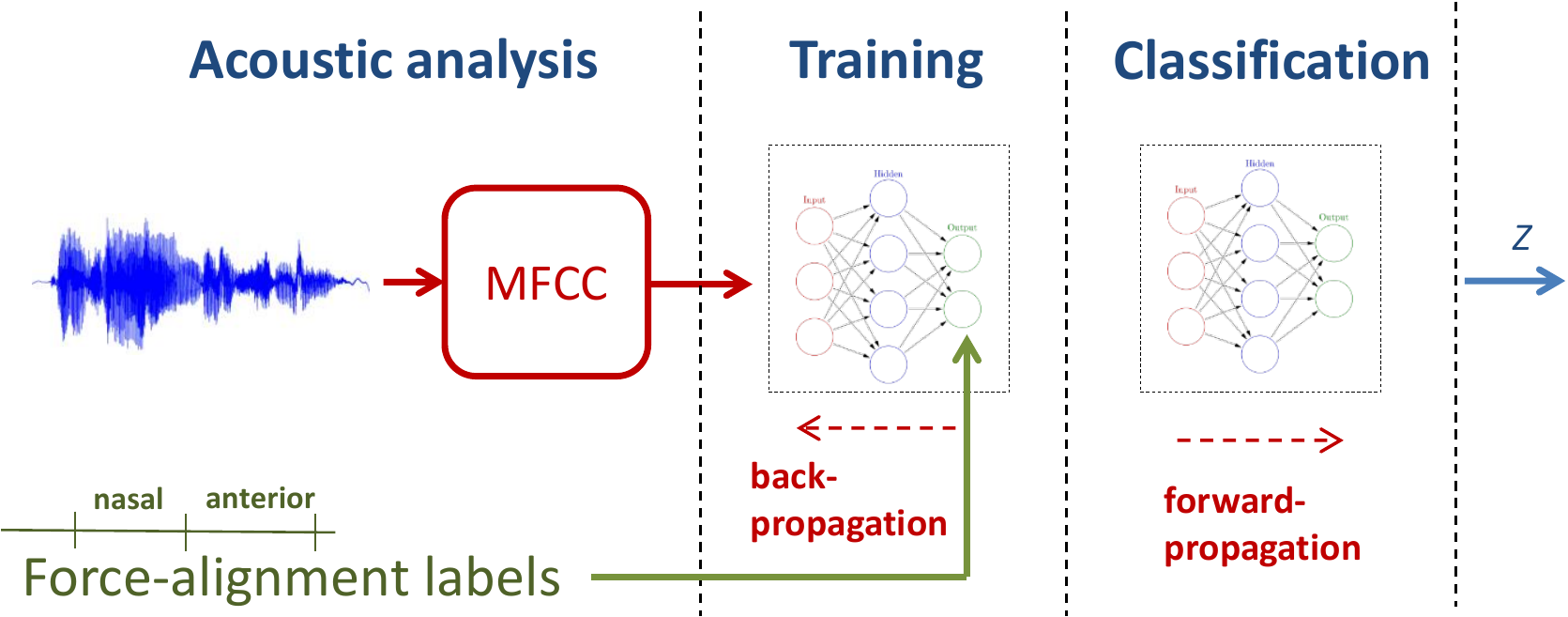} %
\label{fig:adnn}
}
\hfil
\subfloat[b] {
  \includegraphics[width=\linewidth]{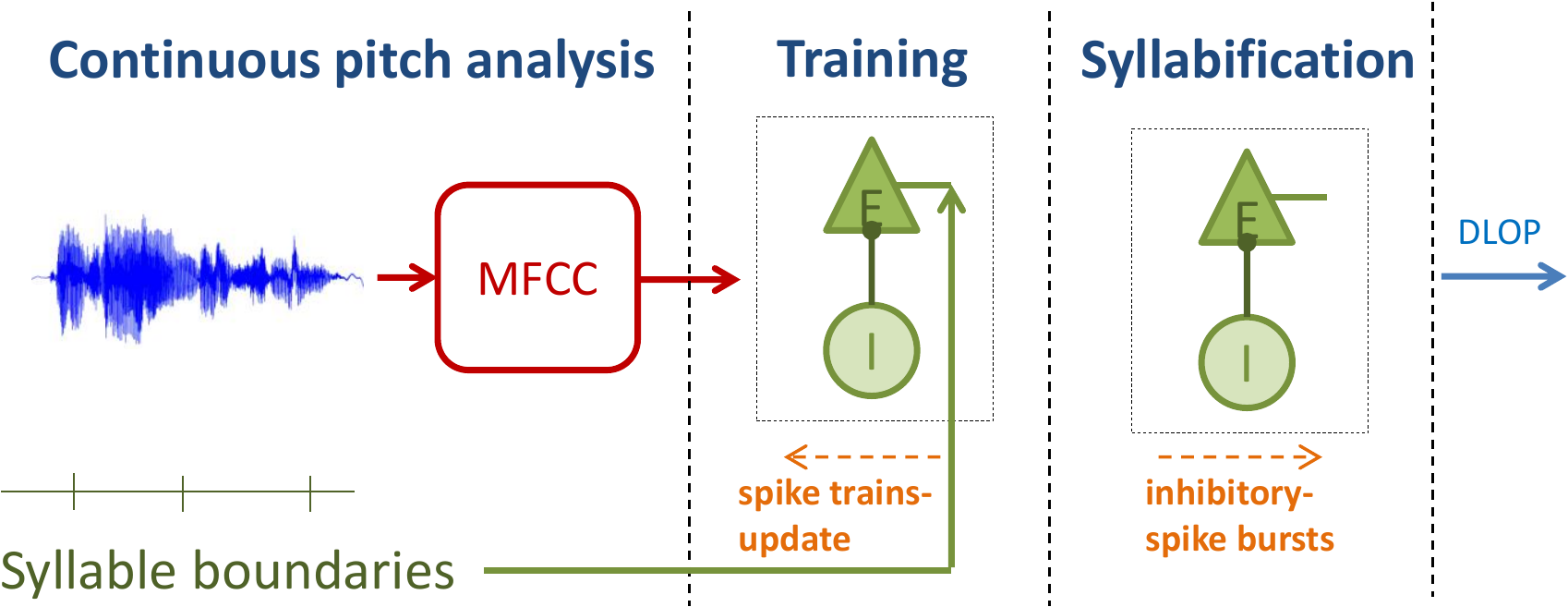} %
\label{fig:sdnn}
}
\hfil
\subfloat[c] {
  \includegraphics[width=\linewidth]{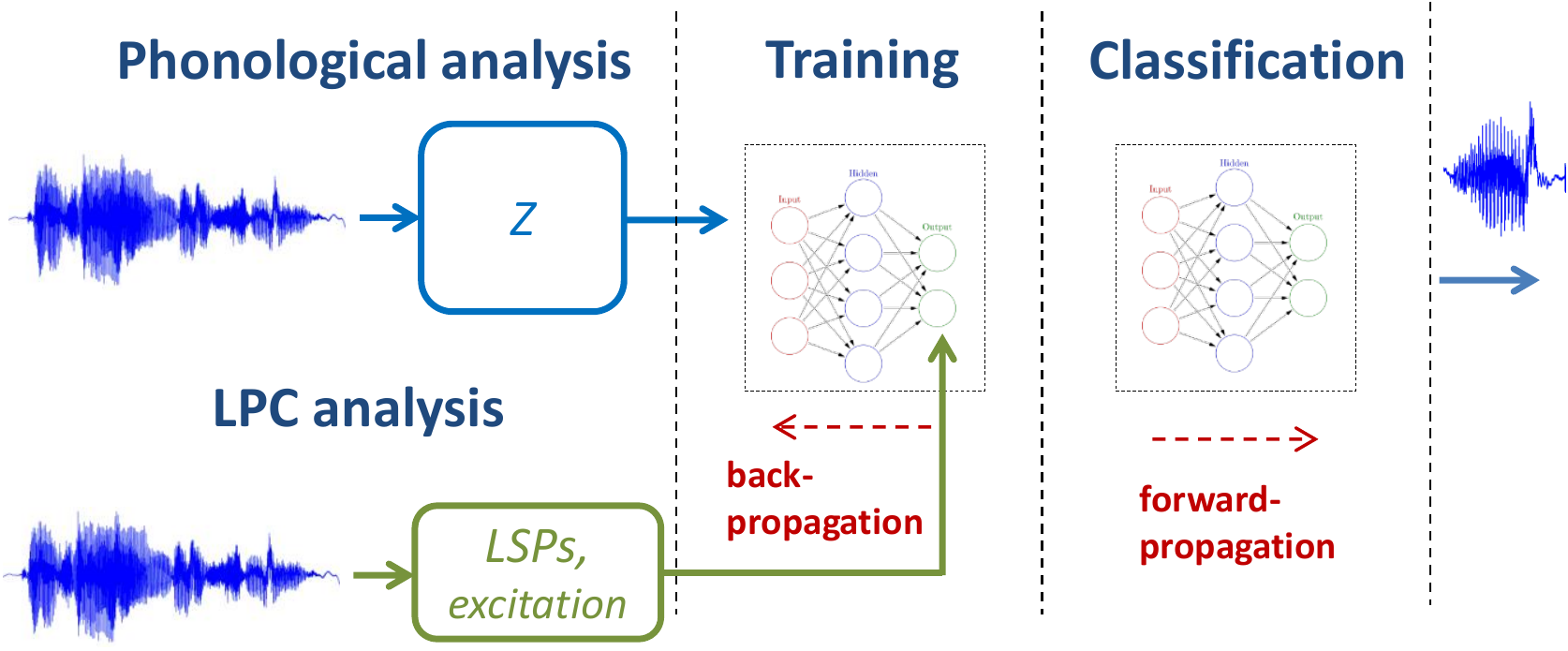} %
\label{fig:snn}
}
\caption{{\rv Training and inference stages for an analysis DNN shown on
  \ref{fig:adnn}, a spiking NN shown on \ref{fig:sdnn}, and a synthesis
  DNN shown on \ref{fig:snn}. Training of the analysis DNN and 
  the spiking NN requires phonetic and syllabic boundary labels,
  respectively, whereas training of the synthesis DNN does not require
force-aligned labels.}}
\label{fig:training}
\end{figure}

{\rv Figure~\ref{fig:training} shows the training and inference stages of the
three different NNs. For phonological analysis DNN training,} we
first aligned the WSJ training data using HMM acoustic models trained
for the baseline system. Then, considering the three different
phonological systems, GP, SPE and eSPE, the three different banks of
DNNs were trained on the 90\% subset of the training set, and the
remaining 10\% subset was used for cross-validation. For all systems,
the labels of phonemes were mapped to the respective phonological
classes. In total, $K$ DNNs (12 for GP, 15 for SPE and 21 for eSPE)
were trained as phonological analyzers using the short segment
(frame) alignment, with two output labels indicating whether the $k$-th
phonological class exists for the aligned phoneme or not. The 
architecture of the DNNs was $351 \times 1024 \times 1024 \times 1024
\times 2$ neurons, determined empirically. The input vectors were 
39-order MFCC features with a temporal context of 9 successive frames.

The training was initialized using deep belief network pre-training
done by the single-step contrastive  divergence (CD-1) procedure of
\cite{Hinton06}. The DNNs with the softmax output function were then
trained using a mini-batch based stochastic gradient descent algorithm
with the cross-entropy cost function of the Kaldi
toolkit~\cite{PoveyASRU2011}. Tables~\ref{tab:GPaccuracies}, \ref{tab:SPEaccuracies}
and \ref{tab:eSPEaccuracies} list the phonological classes and
detection accuracy for GP, SPE and eSPE respectively. The DNN
outputs for individual phonological classes determine the phonological
posterior probabilities.

\begin{table} [h]
  \caption{\label{tab:GPaccuracies} {\small \it Phonological classes
      and classification accuracy of the GP analysis at frame level.}}
\vspace{2mm}
\centerline{
\begin{tabu}{|l|c|c|[0.8pt]l|c|c|}
\hline
Phonolog. & \multicolumn{2}{c|[0.8pt]}{Accuracy (\%)} &
Phonolog. & \multicolumn{2}{c|}{Accuracy (\%)} \\
\cline{2-3} \cline{5-6}
features & train & cv & features & train & cv \\
\hline \hline
A & 95.2 & 93.6 & i & 97.8 & 96.9 \\
a & 98.6 & 98.0 & N & 98.9 & 98.4 \\
E & 95.4 & 93.7 & S & 97.0 & 95.8 \\
H & 97.0 & 95.9 & u & 98.7 & 98.0 \\
h & 97.4 & 96.4 & U & 96.7 & 95.4 \\
I & 96.7 & 95.7 & silence & 99.4 & 99.1  \\
\hline
\end{tabu}}
\end{table}

\begin{table} [h]
  \caption{\label{tab:SPEaccuracies} {\small \it Phonological classes
      and classification accuracy of the SPE analysis at frame level.}}
\vspace{2mm}
\centerline{
\begin{tabu}{|l|c|c|[0.8pt]l|c|c|}
\hline
Phonolog. & \multicolumn{2}{c|[0.8pt]}{Accuracy (\%)} &
Phonolog. & \multicolumn{2}{c|}{Accuracy (\%)} \\
\cline{2-3} \cline{5-6}
features & train & cv & features & train & cv \\
\hline \hline
vocalic & 97.3 & 96.5 & round & 98.7 & 98.1 \\
consonantal & 96.3 & 95.0 & tense & 96.6 & 95.3 \\
high & 97.0 & 95.7 & voice & 96.5 & 95.6 \\
back & 96.2 & 94.8 & continuant & 97.3 & 96.3 \\
low & 98.4 & 97.6 & nasal & 98.9 & 98.4 \\
anterior & 96.8 & 95.6 & strident & 98.7 & 98.2  \\
coronal & 96.1 & 94.6 & rising & 98.6 & 97.8 \\
\hline
\end{tabu}}
\end{table}

\begin{table} [h]
  \caption{\label{tab:eSPEaccuracies} {\small \it Phonological classes
      and classification accuracy of the eSPE analysis at frame level.}}
\vspace{2mm}
\centerline{
\begin{tabu}{|l|c|c|[0.8pt]l|c|c|}
\hline
Phonolog. & \multicolumn{2}{c|[0.8pt]}{Accuracy (\%)} &
Phonolog. & \multicolumn{2}{c|}{Accuracy (\%)} \\
\cline{2-3} \cline{5-6}
features & train & cv & features & train & cv \\
\hline \hline
anterior & 96.7 & 95.5 & low & 98.5 & 97.7 \\
approximant & 98.3 & 97.4 & mid & 96.3 & 95.0 \\
back & 96.5 & 95.1 & nasal & 98.9 & 98.4 \\
continuant & 97.4 & 96.4 & retroflex & 99.1 & 98.7 \\
coronal & 96.9 & 95.7 & round & 97.3 & 96.2 \\
dental & 99.6 & 99.4 & stop & 98.0 & 97.2  \\
fricative & 98.4 & 97.8 & tense & 94.7 & 92.5 \\
glottal & 99.9 & 99.8 & velar & 99.5 & 99.1 \\
high & 96.7 & 95.4 & voiced & 96.8 & 95.8 \\
labial & 98.9 & 98.1 & vowel & 96.5 & 95.3 \\
\hline
\end{tabu}}
\end{table}

\subsubsection{Prosodic analysis SNN}

The prosodic analysis SNN is based on an interconnected network
composed of 10 excitatory and 10 inhibitory leaky integrate-and-fire
neurons. Its principles are based on findings on the role of slow
neural oscillations in the auditory cortex for natural speech parsing
\cite{Hyafil15}. 

The 13-dim cepstral input vectors are first transformed to unidimensional
time series. We use weighting to allow for some channels/frequencies
to have higher importance, e.g. frequencies around formants likely
to provide more information about syllable boundaries because of the
vocalisation process. Syllable boundaries are characterised by local
minima of the weighted signal; these can be generalised to a
convolution of the temporal kernel and the weighted
signal~\cite{Cernak15neuro}.

Training of the parameters of the spiking NN is based on minimising
the syllabic distance of actual syllable boundaries and those 
produced by the convolution, over the \num{1000} sentences of the training
set. The syllabification program tsylb2 \cite{Fisher1996} was used to
convert phonetician-labelled phonemes and phoneme boundaries into
syllables and syllable boundaries.

The output of the spiking NN is stylized using 3-bit quantization of
$2^{nd}$ order DLOP parameters. To create the codebooks, the logarithm of
the continuous pitch of all the syllables of the Nancy training data
was parametrized with DLOP parameters. The values were linearly
spaced between the $\mu - 3 \sigma$ and $\mu + 3 \sigma$ boundaries,
where $\mu$ is the mean and $\sigma$ is the standard deviation of all
the measurements of the DLOP parameters. One codebook was thus created
for the $1^{st}$ order DLOP parameter, and one for the $2^{nd}$ one.

\subsubsection{Phonological synthesis DNNs}
\label{sec:cepstraldnn}

The speech signals from the training and cross-validation sets of the
Nancy database, down-sampled to 16 kHz, framed by \SI{25}{\ms} windows
with the three different frame shifts: 10, 16 and \SI{20}{\ms}, were
used for extracting both DNN input and output features. The input
features, phonological posteriors $\vec{z}_n$, were generated by the
phonological analysers. The temporal
context of 11 successive frames resulted in input features of
$12 \times 11 \times 1 = 132$, $15 \times 11 \times 1 = 165$ and $21
\times 11 \times 1 = 231$ dimensions, for
the GP, SPE and eSPE schemes, respectively. The output features, the
LPC speech parameters, were  extracted by SSP: $\vec{p}_n$ - Line
Spectral Pairs (LSPs) of 24th order plus gain, $\log(\vec{r}_n)$ - a
Harmonic-To-Noise (HNR) ratio, and $\vec{t}_n$, $\log(\vec{m}_n)$ -
two glottal model parameters~\cite{Phil15}, angle $t$ and magnitude
$\log(m)$ of a glottal pole, respectively. Thus, we used static speech
parametrization of 28th order along with its dynamic features,
altogether of 84th order.

Cepstral mean normalisation of the output features was applied before
DNN training. Altogether, 6 DNNs were trained for 3 different
phonological schemes and 3 time shifts. The DNNs were initialised using
$(K*11) \times 1024 \times 1024 \times 1024 \times 1024 \times 84$
pre-training and DNNs with a linear output function were then trained
by Kaldi with  mean square error cost function.

%% file: experiments.tex
In this section, we present an evaluation of the VLBR phonetic and phonological
NN speech coder, and a comparison to the baseline HMM-based system.
We encoded and decoded 1095 utterances of the Nancy database test
set. In following sections, we present results focusing on:
\begin{enumerate}
  \item \textbf{Phonetic NN speech coder}, comparing the phonetic and
    phonological coding in Section~\ref{sec:pp}.
  \item \textbf{Phonological NN speech coder}, quantifying an impact
    of different phonological schemes and frame shifts on speech
    quality and transmission rates in Section~\ref{sec:obj}.
  \item \textbf{HMM vesus NN speech coders}, comparing the VLBR
    HMM-based and NN speech coding in Section~\ref{sec:subj}.
  \item {\rv \textbf{Intelligibility of NN speech coder} in
    Section~\ref{sec:intl}.}
\end{enumerate}

\subsection{Phonetic NN speech coder}
\label{sec:pp}

Segmental information of the NN speech coder may consist of either
phonetic or phonological posteriors (cf. Section~\ref{sec:segmental}),
transmitted frame by frame. Therefore, we started evaluation by
comparison of the speech quality of the phonetic and phonological speech
coding. To measure the impact of the phonetic and  phonological
posteriors only, the F0 encoding was by-passed with the original F0
signals. To achieve VLBR, binary posteriors have to be used, hence we
were interested which of binary phonetic or phonological posteriors
perform better.

We normalised the phonetic and phonological posteriors to the binary
values (the probabilities above 0.5 are normalized to 1 and the
probabilities less than 0.5 are forced to zero) and used Mel Cepstral
Distortion (MCD)~\cite{Kubichek93} between original and encoded speech
samples as an objective metric to compare overall speech
quality. Lower MCD values indicate higher speech quality of the
encoded speech samples. The segmental information in this experiment
consists of \SI{10}{\ms} framed vectors of either phone posterior
probabilities or eSPE phonological-class posterior probabilities.

\begin{table}[ht]
  \caption{\label{tab:MCD} {\small \it Objective quality evaluation of
      continuous and binary parametric phonetic and phonological
      (eSPE) vocoding, {\rv using Mel Cepstral Distortion (MCD). MCD
        is a measure of error-magnitude, whereby smaller is better.}}}
\vspace{2mm}
\centerline{
\begin{tabu}{|l|c|c|c|}
\hline
Type / MCD [dB] & Re-synthesis & Continuous & Binary \\
\hline
\hline
LPC re-synthesis & 4.75 & -- & -- \\
\hline
Phonetic vocoding & -- & 6.11 & 6.70 \\
Phonological vocoding & -- & 6.20 & 6.46 \\
\hline
\end{tabu}}
\end{table}

Table~\ref{tab:MCD} reports the results.  All results are
statistically significant ($p < 0.01$ of a $t$-test). The first row
shows speech distortion caused by the LPC vocoding. The second and
third rows report the distortions of phonetic and phonological
vocoding done on the top of the LPC vocoding. We can conclude that the
majority of the distortion comes from the parametric vocoder; the
distortion of phonetic/phonological vocoding with continuous features
is about 1.4 dB. While the performance of continuous phonetic and
phonological posteriors is similar, feature normalization has higher
negative impact on the binary phonetic posteriors. The reason why
binary phonological posteriors outperform the phonetic ones could be
in the parallel nature of phonological vocoding. Also, maximum a-posteriori
classification of phonological posteriors is far more
accurate than phonetic posteriors. Therefore, we conclude that
phonological posteriors are more suitable for the NN coder, and 
we selected the binary phonological features in further evaluation.


\subsection{Phonological NN speech coder}
\label{sec:obj}

We used the binary phonological posterior features, and
created a codebook from the unique patterns for each phonological
scheme. Linear 3-bit quantized syllable-based F0 parametrization is
used in this experiment. The 3-bit quantization degrades speech
quality by only about 0.1 dB.

\begin{figure}[ht]
\centering
  \includegraphics[width=.9\linewidth]{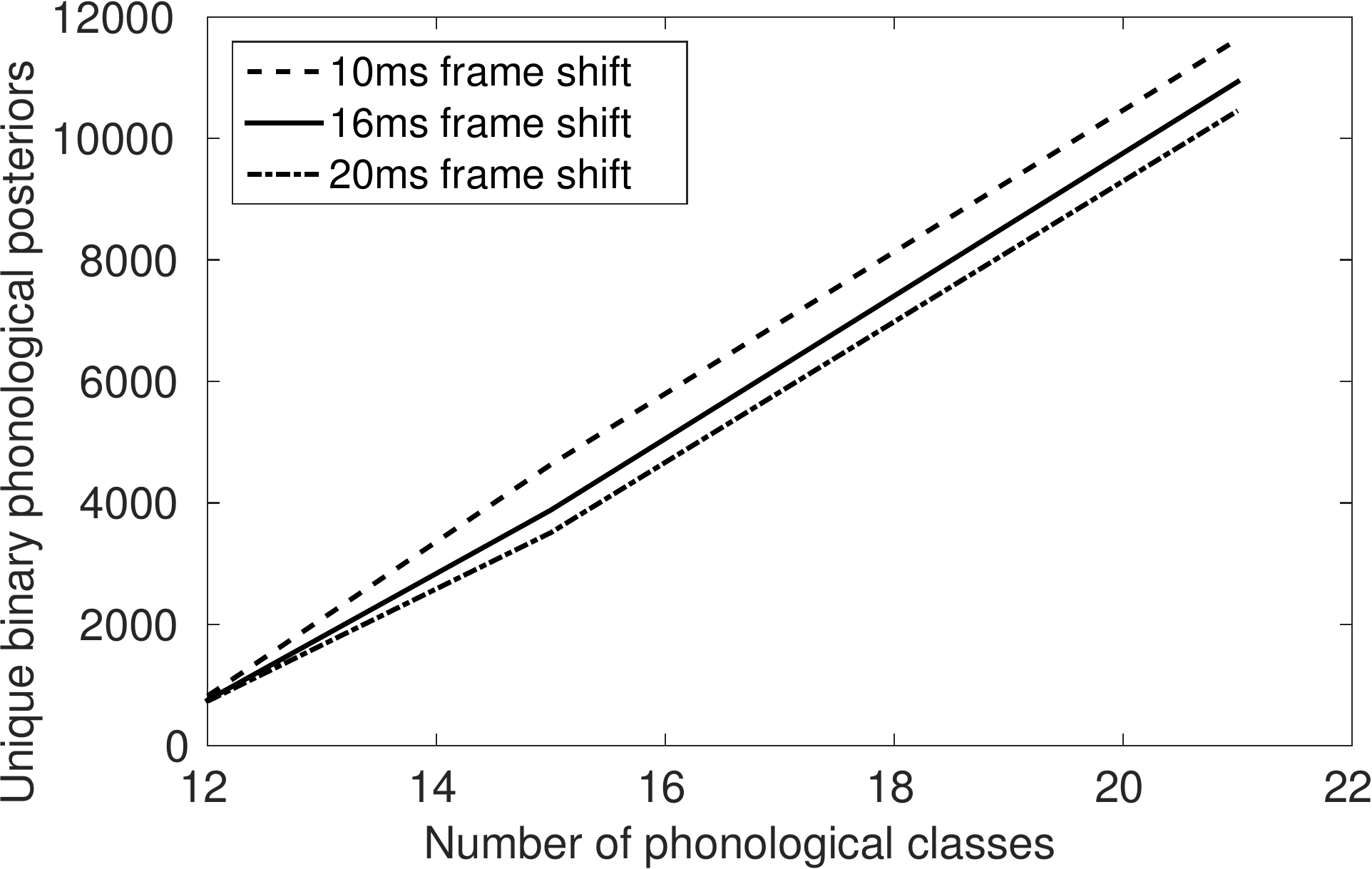} %
  \caption{Linear dependence of unique binary phonological posteriors
    on the number of phonological classes.}
  \label{fig:unique}
\end{figure}

Lower bit-rates can be achieved by using a lower dimensional
phonological scheme. Figure~\ref{fig:unique} shows a linear dependence
of the codebook size on the number of phonological classes. Increasing
the frame shift slightly decreases the number of unique patterns. We
speculate that the number of unique binary phonological posteriors is
related to the number of clustered senones (tied context-dependent HMM
states) used in ASR and TTS acoustic modelling. Then we could
interpret a vector of phonological posteriors as a senone.

Recall that we consider three different phonological systems in
this work. Each system defines a different set of features, with the
dimensions 12, 15 and 21, for the GP, SPE and eSPE phonological
systems respectively. We discuss their meaning and exact definition
in~\cite{Cernak2016a}.

\begin{figure}[ht]
\centering
  \includegraphics[width=.9\linewidth]{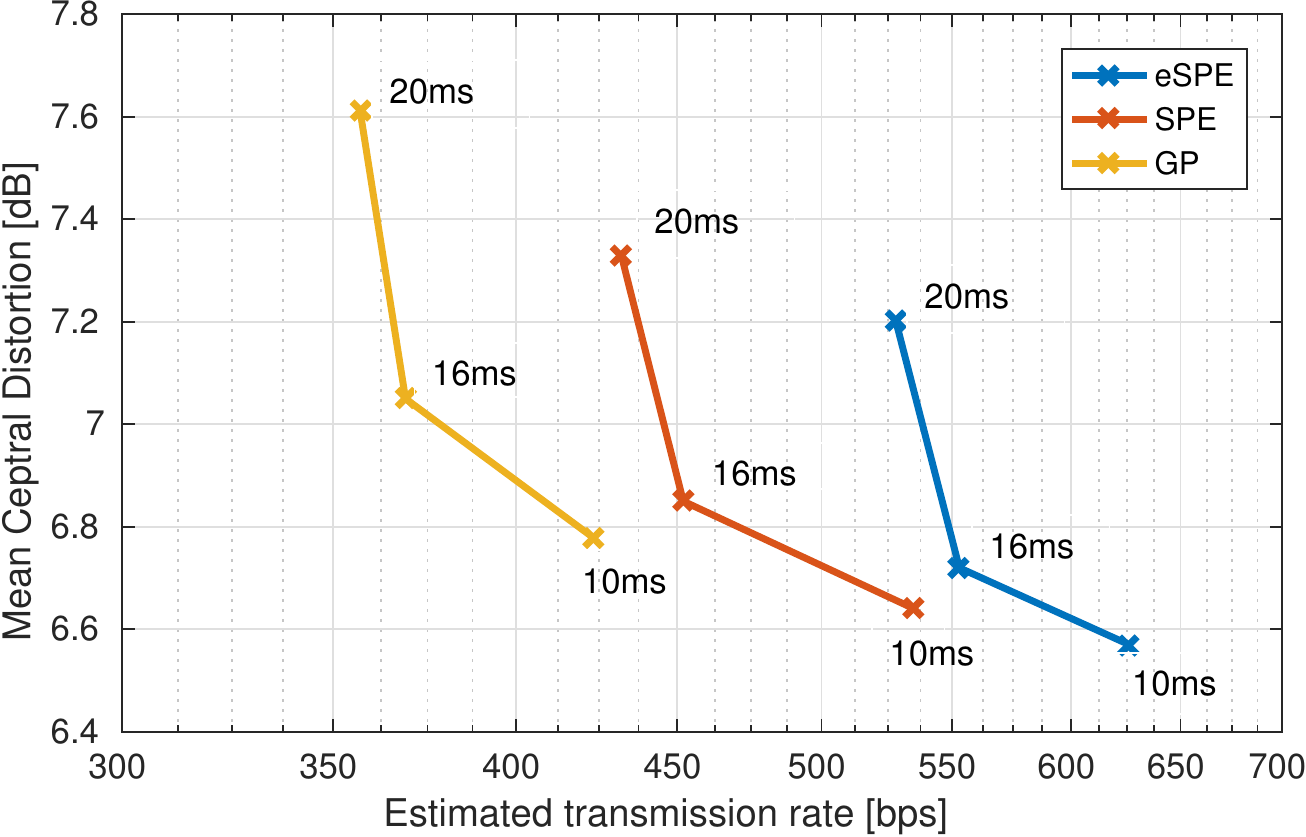} %
  \caption{Estimated transmission rates of the {\rv NN speech
      coding. Data points are estimated for different phonological
      systems and various frame shifts.}}
  \label{fig:bps}
\end{figure}

Figure~\ref{fig:bps} shows the objective evaluation of the NN speech
coding. The transmission rate depends on two variables: the 
phonological scheme that results in smaller or larger
codebooks, and the frame shift. The eSPE codebook consists of
{\rv \num{11632}} unique binary phonological patterns; 14 bits are thus
required to transmit the segmental code. Fewer bits are required for
the SPE and GP codebook, 12 and 10 bits respectively. We can see from
the figure that the difference in quality degradation lies in the
range of about 0.2 dB, therefore we identified the GP system as the
most suitable for the VLBR system. The frame shift has a bigger impact
on the bit rate; for the GP system, when increasing the frame shift from
\SI{10}{\ms} to \SI{16}{\ms}, the degradation increases by 0.3 dB. 
{\rv We consider the frame shift of \SI{16}{\ms} as a good balance between 
the transmission rate and the speech quality.}

\begin{figure}[ht]
\centering
  \includegraphics[width=\linewidth]{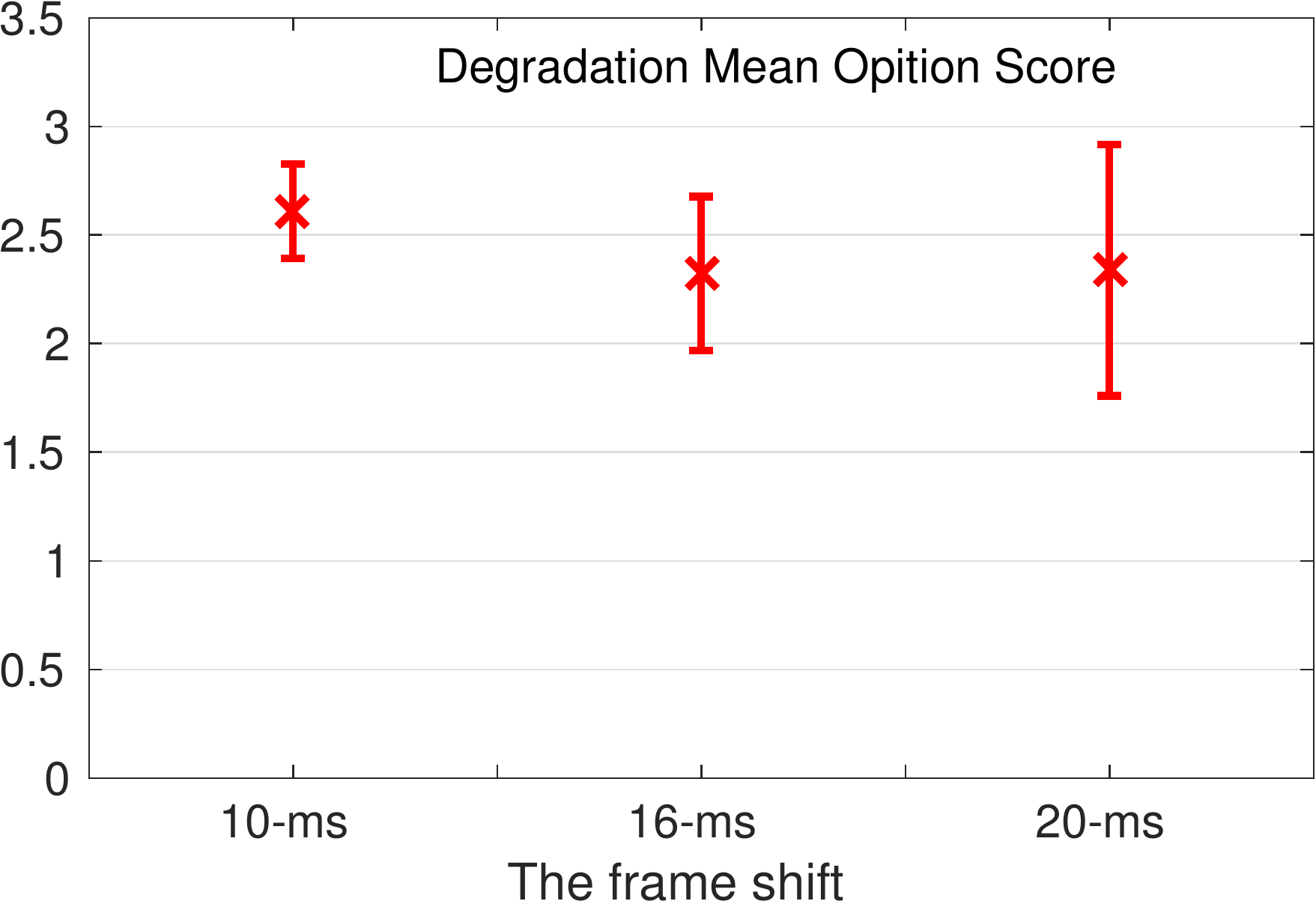} %
  \caption{Subjective evaluation of DNN-based speech coding using the
    GP phonological scheme with various frame shifts. Degradation
    categories are: 3 -- slightly annoying, 2 -- annoying, and 1 --
    very annoying.}
  \label{fig:dmos}
\end{figure}

The overall quality of the NN speech coding was evaluated
subjectively using the Degradation Category Rating (DCR)
procedure~\cite{DMOS} quantifying the Degradation Mean Opinion Score
(DMOS). The aim was to estimate speech encoding quality variations
based on the different frame-shift. The test consisted of 20 randomly
selected utterances from the Nancy test data, at least 2 seconds
long. 37 listeners were asked to rate the degradation of
re-synthesised signals, compared with reference signals, based on their
overall perception. {\rv Figure~\ref{fig:dmos} shows that NN speech
  coding, operating at 360--370 \bps, achieves a MOS of approximately
  2.3}. The figure also shows higher uncertainty of listeners (higher 
standard deviation of MOS) when increasing the frame shift of the DNN
coder. Although similar DMOS is achieved with both the \SI{16}{\ms}
and \SI{20}{\ms} frame shifts, speech coding with the \SI{20}{\ms}
frame shift has much higher standard deviation of subjective
testing. Thus, we selected the GP-based \SI{16}{\ms} frame shift
system as an optimal VLBR NN speech coder for a qualitative and {\rf
  intelligibility} comparison with the HMM coder.

\subsection{HMM versus NN coding}
\label{sec:subj}


The HMM coder creates bursts of segmental errors  when phoneme ASR
fails; this is certainly disruptive to the listeners. It also uses
voiced/unvoiced detection for parametrization of the F0 signal only in
voiced regions. This prosodic encoding plugged-in to the VLBR system
creates additional speech discontinuities and unnatural speech
artefacts. On the other hand, the NN speech coder transmits
information per frame, and parameterizes continuous F0, that
altogether significantly reduces discontinuities during speech
reconstruction at the receiver. Therefore, we continued in perceptual
comparison of the HMM and NN coding.

For this qualitative comparison, we employed a 5-point scale ABX
subjective evaluation listening test~\cite{Grancharov2008}. In this
test, listeners were 
presented with pairs of samples produced by two systems (A and B) and
for each pair they indicated their preference or strong preference 
for A, B, or \emph{both samples sound the same} (X). The material for
the test consisted of 20 pairs of sentences such that one member of
the pair was generated using the GP-based \SI{16}{\ms} frame shift NN
speech coder (system A), and the other member was generated using the
HMM-based speech coder (system B). Random utterances from the test set
of the Nancy database were used to generate the encoded speech.
We chose this GP system as a compromise between the speech quality and
the low bit rate. The subjects for the ABX test were 32 native English 
listeners, roughly equally pooled from experts in speech processing on the one
hand, and completely naive subjects on the other hand. The subjects
were presented with pairs of sentences in a random order with no
indication of which system they represented. They were asked to
listen to these pairs of sentences (as many times as they wanted), and
choose between them in terms of their overall quality. Additionally,
the option X, i.e. \emph{both samples sound the same}, was available
if they had no preference for either of them.

\begin{figure}[ht]
\centering
  \includegraphics[width=\linewidth]{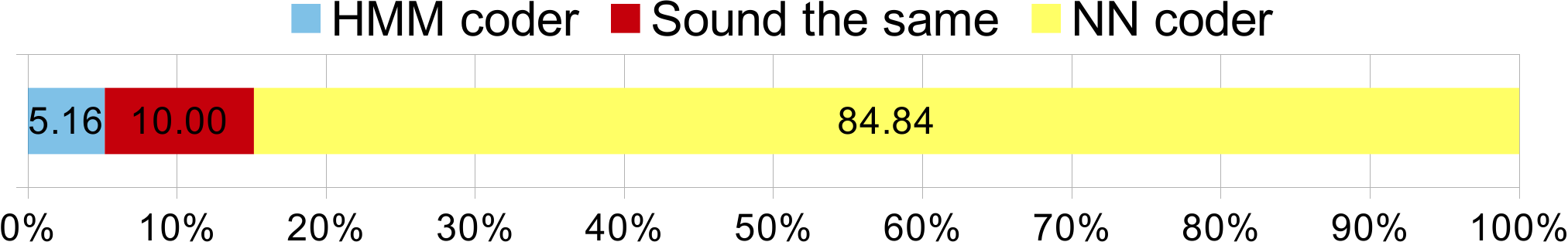} %
  \caption{ABX subjective evaluation test of HMM and DNN/SNN coders.}
  \label{fig:abx}
\end{figure}

As can be seen in Figure \ref{fig:abx}, the NN speech coder
significantly outperforms the HMM-based one. The strong preference and
preference choices of the NN coder achieve 30.62\% and 54.22\% (sum up to
84.84\%) over 0.63\% and 4.53\% (sum up to 5.16\%) respectively for
the HMM-based one. In addition the ``no preference'' choice achieved
10\%.


{\rv
\subsection{Intelligibility of NN coder}
\label{sec:intl}

Finally, we were interested in intelligibility degradation of NN
coding. As a baseline, we selected the Open Source/Free Software
speex codec\footnote{\url{http://www.speex.org/}}. Speex is based on
CELP and is designed to compress voice at bitrates ranging from
\numrange{2.15}{44}~\kbps. It also allows wideband (16 \si{\kHz}
sampling rate) speech compression, the rate with which we designed the
NN coding.

We evaluated {\rf HMM,} NN coding and the Speex coding using the
short-time objective intelligibility (STOI)
measure~\cite{Taal11}. STOI analyzes the short-time magnitudes of
signals within one-third octave bands, and computes the correlations
between reference and test signal band envelopes over segments of
\SI{384}{\ms}. The individual correlation values are then averaged
over bands and segments to yield an objective score in the range [0,
1] that is expected to have a monotonic relation with
intelligibility. This approach was found to provide good predictions
of subjective intelligibility scores for speech coding as used in
cellular networks~\cite{Jorgensen15}. We used the publicly available
MATLAB implementation~\cite{Taal10}.

\begin{table}[ht]
  \caption{\label{tab:STOI} {\small \it The STOI intelligibility
      evaluation of {\rf HMM coding operating at 213 \bps}, NN coding
      operating at 357--626 \bps, and the Speex coding, operating at
      2150 bps.}}
\vspace{2mm}
\centerline{
\begin{tabu}{|l|c|c|c|c|c|}
\hline
Codec  & { \rf HMM}   & NN /        & NN /         & NN /         & Speex \\
type   & {\rf based} &\SI{10}{\ms} & \SI{16}{\ms} & \SI{20}{\ms} & codec \\
\hline
\hline
GP   & \rf{\multirow{4}{*}{0.38}} & 0.70  & 0.68  & 0.64 & \multirow{4}{*}{0.78}\\
\cline{3-5}
SPE  & & 0.71  & 0.70  & 0.66 & \\
\cline{3-5}
eSPE & & 0.71  & 0.71  & 0.67 & \\
\hline
\end{tabu}}
\end{table}

Table~\ref{tab:STOI} shows the intelligibility comparison of {\rf HMM,}
NN and the Speex coding. {\rf We selected the minimal transmission
  rate offered by the Speex coder to compare different coding
  techniques at the most similar bit rates. We can conclude that NN
  coding significantly outperforms the HMM coding in terms of
  intelligibility\footnote{{\rf The lower intelligibility of HMM speech coding
    could be caused by less precise phoneme boundary detection
    performed by the ASR system, which resulted in segmental
    misalignment of the encoded and reference speech samples and that
    impacted the objective intelligibility measure.}} and, compared to
  LPC coding, it degrades intelligibility by only 10\% (for example for
  the GP-based \SI{16}{\ms} frame shift NN speech coder); however, it
  operates at a bit rate approximately 6 times lower.
}

\section{Bit allocation and Complexity}
\label{sec:exp-latency}

The test set contained about {\rv \num{36000} syllables
  (\si{\syllable}) in \SI{5240}{\second}} of speech including
silences. On average, there were {\rv
  \SI{6.8}{\syllable\per\second}}. The transmission code
included the index of the binary pattern along with its duration, the
segmental code, and two indexes of quantized mean and slope of
syllable-based F0 codebooks, along with the syllable duration and the
supra-segmental code. Both segmental and suprasegmental information are
transmitted asynchronously, i.e., the segmental blocks and syllables
have different start, end and duration.

As an example of the bit allocation, Table~\ref{tab:bitrates} shows
the details for the GP system used in the ABX test. Closer analysis
of repeated patterns of the segmental code reveals that the number of
blocks is less than 46\% of the total number of frames and 2 bits is
sufficient to transmit the number of repeated codes. That amounts to
$0.46 \times 56 \times 10 = 257$ {\rv \bps} transmission rate for the
GP codes, and $0.46 \times 56 \times 2 = 52$ {\rv \bps} transmission
rate for the segmental code duration. Because around 10\% of the
transmitted speech is detected as silence (valid for our training
data), we obtain the effective speech frame-rate for the \SI{16}{\ms}
frame shift as $62.5 \times 0.1 = 56$ {\rv \bps}. The transmission
rate of supra-segmental code is constant for all tested NN-based
speech coding systems, as the number of syllables is always
constant. It consists of an index of the F0 mean codebook ($3 \times 6
= 18$ {\rv \bps}), and index of the F0 slope codebook ($3 \times 6 =
18$ {\rv \bps}), and the syllable duration (4 {\rv \bps}). The effective
syllable rate (leading and trailing syllables removed) of our data was 
{\rv \SI{6}{\syllable\per\second}}. The duration of syllables is encoded
by {\rv \SI{4}{\bit}} (covering duration up to \SI{256}{\ms})
that results in encoding supra-segmental information at a constant
60 {\rv \bps}. Altogether, the estimated bit rate for the
NN-based speech coding using the GP phonological scheme and the
\SI{16}{\ms} frame shift is 369 {\rv \bps}.



\begin{table}[!t]
\renewcommand{\arraystretch}{1.3}
\caption{GP-based neural network vocoder bit allocation.}
\label{tab:bitrates}
\centering
\begin{tabular}{|l|c|c|c|}
\hline
Parameter & {\rv \si{\bit\per unit}} & Unit & {\rv \bps}\\
\hline
GP code & 10 & Code block & 257 \\
Duration & 2 & Code block & 52 \\
\hline
F0 mean & 3 & Syllable & 18 \\
F0 slope & 3 & Syllable & 18\\
Duration & 4 & Syllable & 24\\
\hline
Total & & & \textbf{369}\\
\hline
\end{tabular}
\end{table}

The average syllable duration, including leading, trailing and short
pause silences, was \SI{150}{\ms}. As both speech encoding and decoding
processing (forward passes of the NNs) were faster than real-time, we
consider the average syllable duration as an algorithmic latency of
\SI{150}{\ms} of the proposed coder.

According to the G.114, the users
are ``very satisfied'' as long as latency does not exceed
\SI{200}{\ms}~\cite{G114}. {\rv G.114 refers to total latency of
  the complete system including algorithmic- and transmission-latency. For
  example the 3GPP Enhanced Voiced Services codec constrains
  algorithmic latency to \SI{32}{\ms}, such that the expected
  network-latency of around \SI{150}{\ms} can be contained within the
  total \SI{200}{\ms}  limit. We estimate that the total latency of our
  method leaving \SI{130}{\ms} for network-latency should not
  exceed \SI{280}{\ms}, for which G.114 rates users as
  ``satisfied''.}

{\rv 

\subsection{Complexity of NNs}
\label{sec:complexity}

Since most coders are assumed to be used in portable devices, we
evaluated also the suitability of NNs for small footprint systems.

Running the NN speech coder based on the GP system requires 12
analysis DNNs, 1 synthesis DNN and 1 SNN on a device. Since the SNN is
a small footprint system that consists of just 10 excitatory and 10
inhibitory leaky integrate-and-fire neurons, we focus rather on
DNNs. Each analysis DNN was trained with 2.46 million parameters that
would require about \SI{10}{\mega\byte} memory (parameters stored in
single-precision floating-point format). The synthesis DNN was trained
with 3.31 million parameters that would require \SI{13.2}{\mega\byte}
memory. All DNNs thus require \SI{132}{\mega\byte}.

{\rf
The computational complexity of the proposed end-to-end NN speech coder
depends mostly on the size of NNs. The speech encoding and decoding is
realised as a single NN forward pass, and thus its computational
complexity is about $N_w$; that is, the number of weights of the NNs.
}

Recent research on complexity reduction of NNs (for example
\cite{Hinton15,Nakkiran15,Sindhwani15,highwayDNN}) showed a possible 80\%
memory reduction. We then estimate that the reduced memory could fit the
runtime memory of Digital Signal Processors (DSPs) available in typical
smart-phone hardware~\cite{DeepEar}. Effective methods for small footprint
development rely on exploiting the structured representation of weight
matrices~\cite{Sindhwani15}. The low-rank factorization and Toeplitz matrices
are found to enable significant memory reduction with minor or no reduction in
performance~\cite{Nakkiran15,Sindhwani15}. Furthermore, the knowledge in a
large DNN can be distilled in a smaller size DNN if the output of the large DNN
is used as the labels (instead of the hard labels) for training the small size
DNN~\cite{Hinton15}. We plan to explore the idea of soft target training in the
context of DNN based speech coding in our future studies.  }

%% file: conclusions.tex
VLBR speech coding based on a recognition/synthesis paradigm is
normally either corpus-based (using unit-selection approach), or
HMM-based (using HMM-based ASR or TTS). We have designed and presented
a NN-based speech coding composed of deep NNs and a spiking NN; the
solution represents an end-to-end neural network based VLBR speech
coding.

We have compared phonetic and phonological NN coding; given the
binary nature of the phonological posteriors, they outperform
binary phonetic posteriors. Further, we have compared three different
phonological systems, and we conclude that an optimal NN speech
coder can be designed by using the phonological posteriors defined by
the Government Phonology classes. By selecting a frame shift of
\SI{16}{\ms}, the NN coder operates at 369 {\rv \bps} with a latency
of \SI{150}{\ms}.

Listener preference evaluation of HMM and NN-based speech coders
showed that a NN speech coder with continuous F0 modelling is
significantly preferred by (85\% of) the listeners. {\rf We have concluded
that NN coding significantly outperforms the HMM coding in terms of
intelligibility.} As we have used an open-source experimental framework
with a rather standard LPC vocoder, there is potential for higher
speech quality from NN speech coding in future by improved parametric
vocoding. Table~\ref{tab:MCD} shows that more than 77\% of all
degradation comes from the parametric vocoding.

{\rf
The design of the proposed coder is simplified to just the three kinds
of neural networks. Our future work will be focused on investigations
of computational complexity reduction of the NN speech coding, and
further speech quality improvements targetting quality of LPC coding
at 2 \kbps. We believe that this coding approach will then become more
feasible for a range of computation platforms that may be used in
military and tactical communication systems.}